\documentstyle[12pt]{article}
\catcode`\@=11
\@addtoreset{equation}{section}
\def\theequation{\arabic{section}.\arabic{equation}}
\def\section{\@startsection{section}{1}{\z@}{3.5ex plus 1ex minus
   .2ex}{2.3ex plus .2ex}{\bf}}
\def\subsection{\@startsection{subsection}{2}{\z@}{3.25ex plus 1ex minus
   .2ex}{1.5ex plus .2ex}{\bf}}

\def\appendix{ \setcounter{section}{0}\setcounter{equation}{0}
\def\theequation{a.\arabic{equation}}}
\def\anp#1#2#3{
        Ann. Phys. (N.Y.) {\bf #1 }, #2 (19#3)}

\def\phl#1#2#3{
        Phys. Lett. {\bf #1}, #2 (19#3)}
\def\prl#1#2#3{
        Phys. Rev. Lett. {\bf #1}, #2 (19#3)}
\def\rmp#1#2#3{
        Rev. Mod. Phys. {\bf#1}, #2 (19#3)}

\def\prd#1#2#3{
        Phys. Rev. D {\bf #1}, #2 (19#3)}
\def\nup#1#2#3{
        Nucl. Phys. {\bf #1}, #2 (19#3)}
\def\zpc#1#2#3{
        Z. Phys. C {\bf #1}, #2 (19#3)}

\def\beq{\begin{equation}}
\def\eeq{\end{equation}}
\def\mmag{m_{mag}}
\def\sD{\, ^{\textstyle *} \!\! \Delta}
\def\sDi{\, ^{\textstyle *} \!\! \Delta^{-1}}
\def\sDt{\, ^{\textstyle *} \!\! \Delta_{t} }
\def\sDl{\, ^{\textstyle *} \!\! \Delta_{\ell} }
\def\sGa{\, ^{\textstyle *} \! \Gamma }
\def\sF{\, ^{\textstyle *} \!\! F }
\def\sPi{\, ^{\textstyle *} \! \Pi }
\def\sPit{\, ^{\textstyle *} \! \Pi_{3g} }
\def\sPif{\, ^{\textstyle *} \! \Pi_{4g} }
\def\rht{\, ^{\textstyle *} \!\! \rho_{t}}
\def\rhl{\, ^{\textstyle *} \!\! \rho_\ell}
\def\rhlt{\, ^{\textstyle *} \!\! \rho_{\ell,t}}

\def\om{\omega}
\def\hk{\hat{k}}

\def\cc{{\cal C}}

\def\Ps{\not \!\! P}
\def\sf{\Sigma_f}
\def\speq{\;\; = \;\;}
\def\spp{\; + \;}
\def\spm{\; - \;}
\def\intk{\; \int \frac{d^3 k}{(2 \pi)^3} \; }
\def\intw{\; \int^{+\infty}_{-\infty} \; \frac{d \omega}{\omega} \; }
\def\et{E^{\, t}}
\def\el{E^{\, \ell}}
\def\eM{E^M}
\def\dPt{\delta \Pi_t}
\def\dPl{\delta \Pi_\ell}
\begin{document}
\begin{titlepage}
\begin{flushright}
BNL--P--1/92\\
December, 1992\\
\end{flushright}
\vfill
\begin{center}
{\Large \bf Damping rates for moving particles \\
in hot $QCD$}\\
\vfill
{\large \bf Robert D. Pisarski}\\
Department of Physics\\
Brookhaven National Laboratory\\
Upton, New York  11973 \\
\vfill
{\large \bf Abstract}
\end{center}
\begin{quotation}
Using a program of perturbative resummation
I compute the damping rates for fields at nonzero
spatial momentum to leading order in weak coupling in
hot $QCD$.
Sum rules for spectral densities are used to simplify the
calculations.
For massless fields the damping rate has an apparent
logarithmic divergence in the infrared limit, which
is cut off
by the screening of static magnetic
fields (``magnetic mass'').  This demonstrates how
at high temperature even perturbative quantities are sensitive
to nonperturbative phenomenon.
\end{quotation}
\vfill
\end{titlepage}

\section{Introduction}

For an asymptotically free theory such as $QCD$, at high
temperature perturbation theory is a reasonable
first approximation.  Even if initially there are no bare masses,
in an interacting plasma
mass scales small relative to the temperature $T$
are generated radiatively.  With
$g$ the $QCD$ coupling constant,
elementary diagramatic techniques show that quasiparticles
acquire thermal ``masses'' of order $gT$ at one loop order.
For example, both time dependent electric and magnetic fields are
screened by a thermal gluon mass $m_g \sim gT$, as are static electric
fields.  Static magnetic fields are not screened to this order,
since the plasma is one of (colored) electric charge.  It is
expected that static magnetic fields are screened nonperturbatively
by a ``magnetic mass'' [\ref{mag}].

The thermal masses are related to the real part of the pole
in the quasiparticle propagator.  The imaginary part of the pole
is proportional to the damping rate, $\gamma$, and determines
how rapidly a system near equilibrium approaches
it.  Unlike the screening lengths, which are easy to compute,
even to lowest order
in $g$ the damping rates can only be computed consistently
after the resummation of an infinite set of diagrams, termed
hard thermal loops [\ref{rpa}-\ref{ft}].  While infinite,
the entire series
of hard thermal loops can be succintly expressed in terms
of simple effective actions [\ref{gen}].  This resummation program
has been applied to a variety of problems [\ref{bpa}-\ref{bpc},
\ref{bpy}-\ref{bnn}].

The damping rates are inevitably of order $g^2T$, but they
depend in an interesting fashion on how fast the quasiparticle
is moving through the thermal medium [\ref{rpa}].
As a consequence of
Landau damping, resummation produces an effective gluon
that propagates below the light cone, with
damping dominated by scattering off of spacelike effective gluons.
If the incident field
is initially at rest, both the recoil field and the effective gluon
carry nonzero energy and momentum.  While the calculations
are involved, the effective gluon only probes energies and momenta
of order $gT$, and so the damping rate is just some pure number
times $g^2 T$ [\ref{bpb},\ref{bpc},\ref{bpy},\ref{kkm}].

If the initial field is in motion, however,
the effective gluon can be emitted at
ninety degrees relative to the incident (and final) direction.
When the field is moving sufficiently fast, the effective
gluon carries almost zero energy and
spatial momentum, and yet still contributes
to the damping rate.  In this way, transverse effective gluons
probe the static magnetic sector,
and so are sensitive to the presence of a magnetic mass.
For example,
in an $SU(N)$ color gauge theory the damping rate
for a gluon moving with momenta of order $T$ is shown in sec. IV to be
\beq
\gamma_t \; = \; \frac{g^2 N T}{8 \pi} \; \left(
\; ln \left( \frac{m_g^2}{\mmag^2 + 2 \mmag \gamma_t} \right) \; + \;
1.09681... \right) \; ;
\label{e1.1}
\eeq
for technical reasons this expression only holds in the
limit where $\mmag \gg \gamma_t$.

Over large distances the behavior of static magnetic fields is controlled
by a purely bosonic
effective gauge theory in three dimensions, with a (dimensional)
coupling constant $= g^2 T$.  Assuming that bosonic $QCD$ in three
dimensions has a mass gap,
the magnetic mass must be a number times $g^2 T$, $\mmag \sim g^2 T$,
up to a possible
factor of $\sqrt{ln(1/g)}$ [\ref{magc}].
Thus the damping rate in (\ref{e1.1})
is of order $\gamma_t \sim g^2 T \, ln(1/g^2)$.
It was first shown in ref. [\ref{rpa}] that factors of $ln(1/g^2)$
are generic to damping rates at nonzero velocity: here I consider
more carefully how the logarithm is cut off, and try to evaluate
the constant under the logarithm.   Thus while (1.1) holds for
$\mmag \gg \gamma_t$, unless $\mmag$ is some very large number
times $g^2 T$, this limit probably does not apply:
more likely, $\mmag \leq \gamma_t$.  I present this result
to demonstrate that it is possible,
at least in principle, to compute the
constant under the logarithm.

The formula in (1.1) does not apply to hot $QED$.
In hot $QED$ the behavior of static magnetic fields is
determined by
bosonic $QED$ in three dimensions.  This is a free theory,
so $m^{U(1)}_{mag} = 0$, and the damping
rate for fast fermions, analogous to
(1.1), is logarithmically divergent.
(The damping rate for fast photons is not very interesting: it is
finite and of order $e^3 T$.)
I suspect that the damping rate for fast fermions
in hot $QED$ is finite and of order $e^2 T ln(1/e)$, but
this requires separate analysis beyond that presented here.
Amusingly, as suggested originally in ref. [\ref{rpa}],
the calculation in hot $QED$ is more difficult than
that for hot $QCD$, at least when $\mmag \gg \gamma_t$ [\ref{bnn}].

Notice that the damping rate
appears in the argument of the logarithm.
Lebedev and Smilga [\ref{ls}] first pointed out
that it is necessary to include $\gamma_t$ self consistently,
which is how it enters into the right hand side of (1.1).
Their calculations indicated
that $\gamma_t$ alone suffices to cut off the
logarithmic divergence, even if $\mmag = 0$.
Recently, Baier, Nakkagawa, and Niegawa [\ref{bnn}]
argued that while the damping rate must
be included self consistently, that
if $\gamma_t$ is determined from
the position of the singularity in the propagator, then
without the magnetic mass, by itself $\gamma_t$
does not cut off the logarithmic divergence.
My result in (1.1) accords with their arguments.
This happens because the analytic structure of the
propagator is rather complicated, with
an unexpected branch cut appearing off the physical sheet,
near the pole in the quasiparticle propagator.
The pole has an imaginary part $\gamma_t$; the branch
cut begins at a point that is seperated by an amount
$\mmag$ from this pole.  The restriction that
$\mmag \gg \gamma_t$ arises because it is much simpler
to treat the case when the branch cut and the pole are
far from each other than if they are close.

Linde first argued that a magnetic mass
renders the free energy sensitive to nonperturbative effects at
four loop order [\ref{mag}].
The above shows that for fast fields
the damping rates are sensitive
to such effects even at {\it leading} order.
I confess that
while I introduce the magnetic mass in a plausible fashion,
it is at best a
caricature of nonperturbative effects.
Moreover, the appearance of the magnetic mass
does not imply that calculations are
fruitless: perhaps a marriage of
perturbation theory (as in (1.1)) and lattice gauge
theory (to determine $\mmag$) can be arranged.

In sec. II I discuss I derive some necessary sum rules
and therefy introduce the magnetic mass.
In sec. III the damping rate for a slow, heavy fermion is computed.
For kinematic reasons this damping rate does not probe
very small momenta, of order $g^2 T$, and so is insensitive to
the magnetic mass or to the details of analytic continuation.
The damping rate for fast, massless
quarks and gluons is computed in sec. IV.
In sec. V the Ward identities are used to compute
the leading logarithmic dependence in the
damping rates of quarks and gluons traveling with
momenta greater than order $g^2 T$.
An appendix discusses how the the sum rules of sec. II
can be used to compute the term of order $g^3$ in the
free energy.

For calculational ease all damping rates are computed in Coulomb
gauge.  I appeal to general proofs of gauge invariance in
refs. [\ref{bpa}] and [\ref{kkr}] to establish that this result
is independent of the choice of gauge.  Recently, Baier, Kunstatter,
and Schiff [\ref{bks}] observed that naive calculation in covariant
gauges appear to
violate these general proofs.  The dilema was resolved by Rebhan
[\ref{reb}], who demonstrated that an infrared regulator is required
to treat the mass shell singularities which arise in covariant
gauges; see, also, refs.
[\ref{bpc}], [\ref{nnp}], and [\ref{bks}].
I avoid these delicacies by sticking with Coulomb gauge,
but note that explicit calculations in other gauges may well require
the introduction of infrared regulators.

\section{Sum rules}

The conventions and notation of ref. [\ref{bpa}] are followed.
For an $SU(N)$ gauge theory with $N_{f}$ flavors of
massless quarks in the fundamental representation,
the effective gluon mass induced by the thermal medium is
\beq
m_g^2 \speq \left(N \spp \frac{N_f}{2} \right) \;
\frac{g^2 T^2}{9} \; .
\label{e2.0}
\eeq
In Coulomb gauge the only nonzero components of the gluon
propagator $\sD^{\mu \nu}$ are
$\sD^{00}(K) = \sDl(K)$
and $\sD^{ij}(K) = ( \delta ^{ij} - \hk^i \hk^j) \; \sDt(K)$.
Here the gluon four momentum is $K^\mu = (k^0,\vec{k})$
and $\hat{k} = \vec{k}/k$; analytic continuation to real energies,
with $k^0 = i \, \omega$, is implicit.
When the
momentum is soft, with $\omega$ and $k$ of order $m_g$,
the effective plasmon and transverse gluon propagators are
given by the tree term, plus the corresponding hard thermal loop
[\ref{kin},\ref{klwg}]
\beq	\sDl^{-1}(K)  \; = \; k^2
\; - \; 3 \, m_g^2 \; Q_1\left(\frac{ik^0}{k}\right) \; ,
\label{e2.1} \eeq
\beq	\sDt^{-1}(K) \; = \; K^2
\; - \; \frac{3}{5} \; m_g^2  \left( \; Q_3\left(\frac{ik^0}{k}\right) \;
				- \; Q_1\left(\frac{ik^0}{k}\right) \;
				- \; \frac{5}{3} \; \right) \; .
\label{e2.2} \eeq
The $Q_n$ are Legendre functions of the second kind.

The propagator determines the spectral densities of the effective
fields.  For the transverse field,
\beq
\sDt(k^0,k) = \int^{1/T}_{0} d\tau \; e^{ik^0 \tau}
\int^{+\infty}_{-\infty} d\om \; \rht(\om,k) \left(
1 + n(\om) \right) e^{- \om \tau} \; ,
\label{e2.3} \eeq
where $n(\om) = 1/(exp(\om/T)-1)$ is the Bose--Einstein
statistical distribution function.  The
transverse spectral density is given by
\beq
\rht(\om,k) = Im \; \sDt(-i \om + 0^+, k)/\pi \; ,
\label{e2.4}
\eeq
and is a sum of pole and cut terms,
\beq
\rht(\om,k) \;
= \; Z_{t}(k) \; \bigg( \delta(\om - \et_k)
\spp \delta(\om + \et_k) \bigg)
\; + \; \beta_t(\om,k) \; \vartheta(k^2 - \om^2) \; ;
\label{e2.5}
\eeq
$\vartheta(x)$ is the step function,
$\vartheta(x) = 1$ for $x > 0$, $= 0$ for $x < 0$.
The delta functions in (\ref{e2.5})
represent the propagation of
transverse gluons as quasiparticles
with energy $\omega = \et_k$ and residue
$Z_t(k)$.  The spectral density
also includes the contribution of a cut below
the light cone, $|\omega| \leq k$,
with spectral weight $\beta_t(\omega,k)$.
This cut is the result of Landau damping
in a thermal distribution.
The plasmon spectral density, $\rhl(\om,k)$,
is defined similarly from $\sDl(K)$, and determines the plasmon
mass shell $\el_k$, residue $Z_\ell(k)$, and
the plasmon cut, $\beta_\ell(\omega,k)$.
Complete expressions for these quantities
are given in refs. [\ref{kin}] and [\ref{rpb}].
For example, about zero momentum
\beq
\rht(\om,0) \;  \speq
\left(- \; \frac{k^2}{m^2_g} \right) \rhl(\om,k)
\speq \frac{1}{2 m_g} \; \bigg( \delta(\om - m_g)
				\spp \delta(\om + m_g) \bigg) \; .
\label{e2.5a}
\eeq
That is, at rest a transverse gluon and the
plasmon are degenerate in energy, $\et_0 = \el_0 = m_g$.  The
residue for the plasmon is a bit unusual --- it is
proportional to $-1/k^2$ --- but this is innocuous [\ref{rpb}].
The mass shells for transverse
gluons and for the plasmon split away from zero momentum,
and are determined numerically as the solution of
transcendental equations.

For later purposes, note that about zero energy the contribution
of the cut terms to the transverse and plasmon spectral densities are
\beq
\beta_t(\om,k) \;
\sim_{\!_{\!\!\!\!\!\!\!\!\!
\scriptstyle{\omega \rightarrow 0}}} \;
\frac{1}{\pi} \; Im \; \frac{1}{k^2 -
3 \pi m^2_g \omega i/(4k) } \; ,
\label{e2.5b}
\eeq
and
\beq
\beta_\ell(\om,k) \;
\sim_{\!_{\!\!\!\!\!\!\!\!\! \scriptstyle{\omega \rightarrow 0}}} \;
\frac{1}{\pi} \; Im \;
\frac{1}{k^2 + 3 m^2_g + 3 \pi m^2_g \omega i/(2k)} \; .
\label{e2.5c}
\eeq
The imaginary terms in each propagator are both due to Landau damping,
proportional to $m^2_g \omega/k$ as $\omega \rightarrow 0$.
The real terms are given by limits of the propagators at zero
frequency.  Thus for the plasmon term $k^2 + 3 m^2_g$ enters in the
denominator,
with $3 m^2_g$ the static electric mass squared.  For the transverse
term only $k^2$ appears, since at this order
static magnetic fields are not screened.

In sec. III only the expressions in (\ref{e2.5b}) and (\ref{e2.5c})
are required.  In sec.'s IV and V we need integrals of the spectral
densities with respect to powers of $\omega$.
These integrals can be evaluated by means of sum rules
[\ref{rpb},\ref{by},\ref{wel}].

The derivation of sum rules is an elementary exercise in complex
analysis.  While familiar, I go through several examples
in order to emphasize the relevant physics.  The essential point
is to turn the integral over $\om$ into a contour integral in
the plane of complex $k^0$; to avoid confusion I relabel complex
$k^0$ as $z$.  For example,
\beq
\int^{+\infty}_{-\infty} \om \; \rht(\om,k) \; d \om
\speq \frac{1}{2 \pi i} \; \oint_\cc
\; z \; \sDt(z,k) \; dz \; .
\label{e2.6}
\eeq
The contour $\cc$ runs counter clockwise around the imaginary $z$
axis.  Since there are no intervening poles, the contour can then be
deformed into a loop at infinity.  For large $z$, the hard
thermal loop $\delta \Pi(z,k)$ falls off as $m^2_g/z^2$, and
the effective propagator behaves as the bare one,
$\sDt(z,k) \sim 1/z^2$.  Thus (\ref{e2.6}) is the
same as for free field theory,
\beq
\int^{+\infty}_{-\infty} \om \; \rht(\om,k) \; d \om
\speq 1 \; .
\label{e2.7}
\eeq
At zero spatial momentum this sum rule is dominated by the pole
terms in the spectral density, and is easily checked by using
(\ref{e2.5a}).

The sum rule in (\ref{e2.7}) is familiar as a
consequence of the equal time
commutation rules.  It is only valid to lowest order in $g^2$,
when the effective propagator includes just the hard thermal loop.
For example, if the effective propagator included the full gluon
self energy at one loop order, then the right hand side of (\ref{e2.7})
is modified by terms for
the (ultraviolet divergent) wave function renormalization constant
of the gluon.  This is distinct from the finite renormalization constant,
$Z_t(k)$, above.

Two other sum rules are needed for what follows.
One is a relation for
the plasmon spectral density:
$$
\int^{+\infty}_{-\infty}
\frac{d \om}{\om} \; \rhl(\om,k) \speq
- \; \frac{1}{2 \pi i} \; \oint_{\cc - {\cal O}} \; \frac{1}{z} \;
\sDl(z,k) \; dz $$
\beq
\speq - \; \frac{1}{k^2} \; + \; \sDl(0,k) \speq
- \; \frac{1}{k^2} \; + \; \frac{1}{k^2 + 3 m^2_g} \; .
\label{e2.9}
\eeq
The contour in the complex plane is now $\cc - {\cal O}$, where
${\cal O}$ is a circle about the origin.  The modification of
contour is required because of the factor of $1/z$ in the integrand:
this factor generates a pole in $z$ whose contribution must be included.
The result on the right hand side is a sum of two terms.  The first
results from deforming $\cc$ into the circle at infinity.
As for the transverse density, the contribution from the hard thermal
loop vanishes at large $z$, and so the integral over $\cc$ gives
the same result as in free field theory, $-1/k^2$.  Secondly, there
is the contribution from ${\cal O}$; there the residue of the integrand
at $z=0$ is just the value of the plasmon propagator at zero frequency,
$\sDl(0,k) = 1/(k^2 + 3 m^2_g)$.  About zero momentum this sum
rule is dominated by the pole terms, (\ref{e2.5a}).

The last sum rule required is superficially similar to that
for the plasmon density:
$$
\int^{+\infty}_{-\infty}
\frac{d \om}{\om} \; \rht(\om,k) \speq
- \; \frac{1}{2 \pi i} \; \oint_{\cc - {\cal O}} \; \frac{1}{z} \;
\sDt(z,k) \; dz
$$
\beq
\speq
\sDt(0,k) \;\; \equiv \;\; \frac{1}{k^2 + \mmag^2} \; .
\label{e2.10}
\eeq
The contour at infinity does not contribute because
$\sDt(z,k)$ falls off as $1/z^2$ at large $z$.
For the contour about the origin, ${\cal O}$, the residue of the
integrand is equal to the value of the transverse propagator at
zero frequency, $\sDt(0,k)$.

Using the effective propagator of (\ref{e2.2}),
which includes just the hard thermal loop,
(\ref{e2.10}) equals $\sDt(0,k) = 1/k^2$.
Unlike the two previous sum rules, about zero spatial
momentum (\ref{e2.10}) is dominated
not by the pole term, (\ref{e2.5a}), but
by the cut term in the spectral density, (\ref{e2.5b}) [\ref{rpb}].

In (\ref{e2.10}) I extend this relation, and introduce the magnetic mass,
by {\it defining} the magnetic mass as the position of
a presumed pole in the static transverse propagator,
$\sDt(0,k) = 1/(k^2 + \mmag^2)$.
This is merely a crude parametrization of the complicated
physics which is responsible for the dynamical generation of a finite
correlation length for static magnetic
fields.  As written, the magnetic mass represents the effects of a single
glueball; surely there is a entire tower of glueball states, none of
which need show up simply as a pole in the transverse propagator.
A better
approach would be to relate the quantities which enter into the damping
rate to the vacuum expectation values of gauge invariant operators, which
could then be computed by lattice gauge theory.
At present this noble goal is beyond my means.

In the calculations of the hard damping rate,
kinematics typically restricts
the integral over the gluon spectral densitites to lie below the light
cone.  For example,
for the sum rule of (\ref{e2.10}),
\beq
\int^{+k}_{-k}
\frac{d \om}{\om} \; \rht(\om,k) \speq
\frac{1}{k^2 + \mmag^2}  \;-\;  \frac{ 2 \, Z_t(k)}{\et_k } \; .
\label{e2.8}
\eeq
That is, the sum rule allows one to exchange an
integral over the cut in the
spectral density, $\beta_t(\om,k)$, for a function of
$\et_k$ and $Z_t(k)$.

To incorporate $\mmag$, the limiting form of
the spectral density in (\ref{e2.5b}) becomes
\beq
\beta_t(\om,k) \;
\rightarrow_{\!_{\!\!\!\!\!\!\!\!\!
\scriptstyle{\omega \rightarrow 0}}} \;
\frac{1}{\pi} \; Im \; \frac{1}{k^2 + \mmag^2 -
3 \pi m^2_g \omega i/(4k) } \; ,
\label{e2.5bb}
\eeq
Using just (\ref{e2.5bb}), the integral
$\int^{+k}_{-k} d\omega \rho_t(\omega,k)/\omega
\sim 1/(k^2 + \mmag^2)$.
For $k \sim \mmag$ this is the dominant term on the
right hand side of (\ref{e2.8}); the pole term
contributes $\sim 1/m^2_g$, which is smaller by order $g^2$.
This shows how the sum rule in (\ref{e2.10}) is dominated
by the cut term at small momenta.
When $k \gg \mmag$ the magnetic mass is negligible,
and we recover the full sum rule.

\section{Damping rate for a slow, heavy fermion}

For a heavy fermion of momentum $P$ and
mass $M$, the bare inverse propagator is
$\Delta_f^{-1}(P) = - i \Ps + M$.  To leading order the fermion
self energy is
\beq
\sf(P) \speq - \; g^2 C_f \; tr \; \left( \gamma^\mu \; \sD_f (P-K)
\gamma^\nu \; \sD^{\mu \nu}(K) \right) \; .
\label{e3.1}
\eeq
For a fermion in the fundamental representation the Casimir constant
$C_f = (N^2-1)/(2N)$; $tr$ represents the integral over the loop four
momentum $K$.  In (\ref{e3.1}) I have replaced the bare propagator by
an effective propagator, $\sD_f$.  This is defined as follows.  For
the bare propagator, the spectral density is
\beq
\rho_f(\omega,k) \speq \left(- \, \omega \gamma^0 \spp i \vec{k} \cdot
\vec{\gamma} \spp M\right) \; \frac{1}{2 \omega}
\left( \delta(\omega - \eM_k)
\spm \delta(\omega + \eM_k) \right) \; ,
\label{e3.2}
\eeq
where $\eM_k = \sqrt{k^2 + M^2}$ is the fermion energy.  To include the
effects of damping I replace the sharp delta function in the spectral
density by a Breit--Wigner form, with width $\gamma_f$:
\beq
\rho_f(\omega,k) =
\left(- \, \omega_f \gamma^0 + i \vec{k} \cdot
\vec{\gamma} + M \right) \;
\frac{\gamma_f}{2 \pi \omega_f} \;
\left( \frac{1}{(\omega_f - \eM_k)^2 + \gamma_f^2}
\spm \frac{1}{(\omega_f + \eM_k)^2 + \gamma_f^2} \right) \; .
\label{e3.3}
\eeq
By the properties of the delta function this reduces to (\ref{e3.2})
as $\gamma_f \rightarrow 0$.
The damping rate $\gamma_f$ turns out to be of order $g^2 T$;
this is $g$ times the natural scale for the gluon spectral
densities, which is set by the thermal gluon mass, $m_g \sim gT$.
This inclusion of higher order effects in a hard propagator
extends the program of resummation outlined in ref. [\ref{bpa}].
I discuss later why it is valid to include these higher
order effects, and not others, after computing $\gamma_f$.
Lebedev and Smilga [\ref{ls}] were the first to introduce the
damping rate in this way.

To simplify the computations I
assume that the particle's motion is nonrelativistic, with a velocity
$v = p/M \ll 1$.  The case of relativistic motion is treated following
the analysis of sec. IV.
In (\ref{e3.1}) the Saclay method
[\ref{bpa}] is used to perform the sum over
$k^0$.  The damping rate is proportional to the the imaginary part
of the self energy on the mass shell.  This
is a sum of two terms, from the plasmon and transverse spectral densities:
\beq
Disc \, \Sigma_f(i \eM_p,p) \speq
2 \, i \, a_t \, (\gamma^0 \spm 1 )
\spm i \, a_\ell \, (\gamma^0 + 1) \; ,
\label{e3.4}
\eeq
where
\beq
a_{\ell,t} \speq \frac{g^2 \pi C_f T}{2} \; \intk
\intw \int^{+\infty}_{-\infty} d\omega_f \;
\frac{\gamma_f}{\omega_f^2 + \gamma_f^2} \; \rhlt(\omega,k)
\; \delta(\omega + \omega_f - \eM_p + \eM_{p-k}) \; .
\label{e3.5}
\eeq
Several approximations have been made to reach
(\ref{e3.4}) and (\ref{e3.5}).
The term of leading order in $g$ is given by replacing
the Bose--Einstein
statistical distribution function for the gluon by
$n(\omega) \sim T/\omega$.
Further, in most instances the fermion spectral parameter,
$\omega_f$, can be replaced
by its average value, equal to the energy on the mass shell.
At nonrelativistic velocities this energy is
just the mass: $\omega_f \sim \eM_{p-k} \sim M$.
The only instance where this is not allowed
is in energy denominators: there,
after taking the discontinuity of the self energy, one of the energy
denominators produces
the delta function for energy conservation in (\ref{e3.5}).  (The other
energy denominators don't contribute, since the spectral density for
the soft gluon only has support for $\omega$ and $k$ of order $m_g$.)
For that term the spectral parameter $\omega_f$ is redefined as
$\omega_f \rightarrow \omega_f + \eM_{p-k}$.

Including the self energy, the renormalized fermion propagator is
$-i \Ps + M - \Sigma_f$.  The pole
in the renormalized propagator is shifted from the bare mass shell
to $(\eM_p + i \gamma_f,\vec{p})$, where $\gamma_f$ is the damping rate,
\beq
\gamma_f \speq - 2 \, a_\ell \spp v^2 \, a_t  \; .
\label{e3.6}
\eeq
For simplicity, assume that the velocity, while small, is
larger than $g$, $1 \gg v \gg g$, so that $\eM_{p-k} \sim M
+ (\vec{p}-\vec{k})^2/(2M) \sim M - p k cos\theta/M$.
The delta function for energy conservation is used to fix the
angle between $\vec{p}$ and $\vec{k}$,
$cos\theta = \hat{p} \cdot \hat{k} = (\omega + \omega_f)/(vk)$,
\beq
a_{\ell,t} \speq \frac{g^2 C_f T}{8 v \pi^2} \;
\int^{\infty}_0 \; k^2 dk
\intw \int^{+\infty}_{-\infty} d\omega_f \;
\frac{\gamma_f}{\omega_f^2 + \gamma_f^2} \; \rhlt(\omega,k)
\; \vartheta \left( vk - |\omega + \omega_f| \right) \; .
\label{e3.7}
\eeq

The step function $\vartheta$ enters to ensure that
$|cos\theta| \leq 1$.
I assume that the constraint is
satisfied separately by
$|\omega| \leq vk$ and $|\omega_f| \leq vk$,
so that the
integrals over $\omega_f$ and $\omega$ decouple; this is
justified following (\ref{e3.17}).
The integral over $\omega_f$ gives
\beq
\int^{v k}_{- \, v k} \; d\omega_f \;
\frac{\gamma_f}{\omega_f^2 +\gamma_f^2} \speq
2 \; tan^{-1} \left( \frac{vk}{\gamma_f} \right) \; .
\label{e3.8}
\eeq

The integral over the plasmon spectral density in $a_\ell$ is
\beq
\int^{v k}_{- \, v k} \; \frac{d\omega}{\omega} \; \rhl(\omega,k)
\; \sim \; - \; \frac{3 m^2_g v}{(k^2 + 3 m^2_g)^2} \; .
\label{e3.9}
\eeq
Since $vk$ in (\ref{e3.9}) is small relative to $k$,
the integral in (\ref{e3.9}) is just $2 v k$ times
the limit of $\rhl(\omega,k)/\omega$ as $\omega \rightarrow 0$,
(\ref{e2.5c}).
The remaining integral over $k$ is finite, dominated by
momenta $k \sim m_g \sim gT$.  Then I can take
$\gamma_f \sim 0$ in (\ref{e3.8}), so that
\beq
a_\ell \speq - \; \frac{g^2 C_f T}{16 \pi} \; .
\label{e3.10}
\eeq
This term is negative, and so from (\ref{e3.6}) a positive
contribution to the damping rate.  While various assumptions were made
about the velocity to obtain (\ref{e3.10}), it is not difficult
to go back to (\ref{e3.5}) and show that one obtains identically the
same result even for a field at rest, (5) of ref. [\ref{rpa}].

For the contribution of the transverse gluons, again
since the integral runs only from $-v k$ to $v k$,
for small velocities only the limiting
form in (\ref{e2.5b}) is required; the integral over it
gives
\beq
\int^{v k}_{ - \, v k} \; \frac{d\omega}{\omega}
\; \rht(\omega,k) \speq
\frac{2}{\pi k^2} \; tan^{-1}
\left( \frac{3 \pi v m^2_g}{4 k^2} \right) \; .
\label{e3.11}
\eeq
The magnetic mass, as enters in (\ref{e2.5bb}), has been
neglected; see the discussion following (\ref{e3.17}).
Corrections to (\ref{e3.11}) are proportional to the velocity.
After rescaling $k \rightarrow \gamma_f k/v$,
(\ref{e3.5}) becomes
\beq
a_t \speq \frac{g^2 C_f T}{2 \pi^3 v} \;
\int^{\infty}_0 \; \frac{dk}{k}
\; tan^{-1}(k) \; tan^{-1}\left( \frac{c^2}{k^2} \right) \; ,
\label{e3.12}
\eeq
where $c$ is a pure number,
\beq
c^2 \speq \frac{3 \pi^2 m^2_g v^3}{4 \gamma_f^2} \; .
\label{e3.13}
\eeq
In this instance it is necessary to keep $\gamma_f \neq 0$,
as otherwise (\ref{e3.12}) develops a logarithmic divergence.
Assume that the velocity lies in the range
\beq
1 \; \gg \; v \; \gg g^{2/3} \; .
\label{e3.14}
\eeq
Then the parameter $c$ is large, since with $m_g \sim gT$ and
$\gamma_f \sim g^2 T$,
$c^2 \sim m^2_g v^3/\gamma_f^2 \sim v^3/g^2 \gg 1$.  It is then
straightforward to compute the integral in (\ref{e3.12}).  The
integrand behaves like $1/k$ only for $c \gg k \gg 1$ and is
otherwise well behaved.  Up to corrections of order $ln(c)/c$,
\beq
a_t \speq \frac{g^2 C_f T}{16 \pi v} \; ln(c^2) \; .
\label{e3.15}
\eeq

Altogether, (\ref{e3.6}), (\ref{e3.10}), (\ref{e3.13}), and
(\ref{e3.15}) give
\beq
\gamma_f \speq \frac{g^2 C_f T}{8 \pi} \left(
1 \spp \frac{v}{2} \;
ln\left( \frac{3 \pi m^2_g v^3}{4 \gamma^2_f} \right) \right) \; .
\label{e3.16}
\eeq
At small velocities, inside the logarithm I can replace
$\gamma_f$ by its value at zero velocity to obtain
\beq
\gamma_f \speq \frac{g^2 C_f T}{8 \pi} \left(
1 \spp \frac{v}{2} \;
ln\left( \frac{16 \pi^3 (N + N_f/2)}{3 C_f^2} \;
\frac{v^3}{g^2} \right) \right) \spp \ldots \; .
\label{e3.17}
\eeq
The coefficient of the logarithm agrees with previous results
[\ref{rpa}, \ref{bnn}].

I now justify separating the constraints on $\omega$
and $\omega_f$ in (\ref{e3.7}).  For the term involving the
longitudinal spectral density there is no question;
$\gamma_f$ can be sent to zero at the outset.
For the transverse spectral density,
the integral over $\omega$ is dominated by $\omega \sim
v k$; from the right hand side of (\ref{e3.11}), the relevant
scale of momenta is $k \sim \sqrt{v} m_g$,
so the dominant
frequences are $\omega \sim v^{3/2} g T$.
For $v \gg g^{2/3}$, then, $\omega \gg g^2 T$, and so
the scale for $\omega$ is much greater than that for
$\omega_f$, which is $\omega_f \sim \gamma_f \sim g^2 T$.
This separation in scales produces the
logarithm in (\ref{e3.16}) and (\ref{e3.17}), and justifies
treating the constraints separately.  This is not
allowed if (\ref{e3.14}) is not obeyed.  For smaller
velocities, $v \leq g^{2/3}$,
the scales in $\omega$ and $\omega_f$ do mix;
ultimately there will be no logarithm, with the term from the
transverse spectral density vanishing
smoothly as $v \rightarrow 0$.  Larger
velocities, $v \gg 1$, puts us in the relativistic regime,
which is the subject of the next section.

Also, since the dominant momenta for the transverse density
are $k \sim \sqrt{v} m_g \gg g^{1/3} m_g \sim \mmag/g^{2/3}$,
it is permissible to neglect the magnetic mass, taking
(\ref{e2.5b}) for the limiting form of the spectral
density instead of (\ref{e2.5bb}).

I have been somewhat careless on one other
point.  The correct damping rate
is given by evaluating the imaginary part of the self energy at
the position of the pole in the propagator.  Including damping, this
pole is off the physical sheet, at $\omega_{pole}
= E_p^M + i \gamma_f$.  Instead,
I evaluated the imaginary part at $\omega = E_p^M$, and assumed
that the continuation to $\omega_{pole}$
is trivial.  Baier, Nakkagawa, and
Niegawa [\ref{bnn}] have recently argued
this continuation can produce
a nonzero contribution to the damping rate.

For velocities which satisfy (\ref{e3.14}), though, these subtleties
can be overlooked.
Suppose I were to evaluate the self energy
not just for $\omega = E_p^M$, but for
for $\omega = E_p^M + \delta E$, with $\delta E$ of order $g^2 T$.
This alters energy conservation, so that in (\ref{e3.7})
$|\omega|$ and $|\omega_f|$ must be $\leq vk +
\delta E$.  For $\delta E \sim g^2 T$, however, this change is
negligible, since $vk \sim v^{3/2} g T \gg g^2 T$ if
$v \gg g^{2/3}$.  Thus the continuation from $\delta E = 0$
to $\delta E = i \gamma_f$ does not affect the result for the damping
to leading order in $g$.

Consequently, for a slow, heavy fermion,
due to kinematic reasons there is
no sensitivity to the magnetic
mass (contrary to what I claimed in ref. [\ref{rpa}]) nor to
details of analytic continuation (unlike ref. [\ref{bnn}], at least
for velocities as in (\ref{e3.14})).
Both of these effects do enter
for a field moving at relativistic velocities.

Why is it that corrections from the imaginary part of the self energy
must be included, and not those from the real part?
While there are
certainly corrections to the mass shell of order
$g^2 T$, at hard momenta these are independent of the
spatial momentum.
By energy conservation in
(\ref{e3.5}), however, at this order
all that enters into the damping rate is the difference
in energies, $\eM_p - \eM_{p-k}$ --- so a constant shift in the
mass shell cancels out.
Similarly, consider the contribution of
those higher loop diagrams
which can be represented as vertex corrections to the diagram
at lowest order.  Assuming that the vertex correction is
fixed by the appropriate Ward identity, if the self energies
are slowly varying functions of momenta, the vertex corrections
are even better behaved, and so will
not generate the type of logarithmic divergences found
above.  (These arguments do not
apply at soft momenta, where the self energies, and so
the vertex corrections, depend nontrivially on momenta.)
Nevertheless, it would
be well worth checking these naive arguments by explicit
calculation at two loop order.

\section{Damping rates for fast fields}

Since it enters accompanied by an overall factor of the velocity,
the logarithmic sensitivity found for the damping rate of
a slow, heavy field is relatively innocuous.
The damping rate of a fast particle is much more sensitive to
small momenta.  I first consider the case of a
transverse gluon at ``hard''
momenta, $p \sim T$.

For a transverse gluon the bare propagator is $\Delta_t^{-1}(P)=
(p^0)^2 + p^2$.
About the mass shell $\et_p = p$,
the bare propagator
behaves as $\Delta_t^{-1} \sim - Z_t^{-1}(p) (\omega - \et_p)$, where
$Z_t(p) = 1/(2 p)$ is the residue for a hard field.
Following Baier, Nakkagawa, and Niegawa [\ref{bnn}],
the damping rate is determined from the discontinuity of the self
energy at the position of the singularity in the propagator.
To take this into account,
I evaluate the discontinuity not just at, but near the
mass shell, by introducing the function
\beq
\Gamma_t(\delta E)
\speq Z_t(p) \; Disc \, \Pi_t(i (\et_p + \delta E), p) \; .
\label{e4.1}
\eeq
Here $\Pi_t$ is the transverse part of the gluon self energy,
$\Pi^{i j}(P) = (\delta^{i j} \spm
\hat{p}^i \hat{p}^j ) \Pi_t(P) + \ldots$.
I first evaluate the function $\Gamma_t(\delta E)$ for
real $\delta E$, and then analytically continue
in $\delta E$ to determine the damping rate from
$\gamma_t = \Gamma_t(i \gamma_t)$.
Implicitly, this definition assumes that the propagator has a true pole
at $\omega = E^t_p + i \gamma_t$,
which I shall show is justified in hot $QCD$ when $\mmag \neq 0$.
In hot $QED$, it is necessary to define the damping rate by the
position of the singularity in the propagator; the singularity
will not just be a simple pole.

By the resummation proceedure of
ref. [\ref{bpa}], to determine the damping rate at hard momentum
requires the evaluation of the self energy which differs only
slightly from the usual one loop diagram.  Kinematics requires that
only one line in the loop is soft, so bare vertices can be used.
With bare vertices,
the only contribution to the discontinuity is from the diagram with
three gluon vertices.  (The quark loop does not contribute to
this order because there is no enhancement from Bose--Einstein
statistics.)  After projecting out the transverse piece of the
gluon self energy,
\beq
\Gamma_t(\delta E) \speq g^2 N \; Z_t^{-1}(p)\; tr \;
\left(
(1 \spm (\hat{k} \cdot \hat{p}) )^2 \; \sDt(K) \spm \sDl(K) \right)
\Delta_t(P-K)   \; .
\label{e4.2}
\eeq
For the soft field, with spectral parameter $\omega$,
the spectral densities are those of sec. II.
For the hard field, with spectral parameter $\omega_h$,
the bare spectral density is replaced by
\beq
\rho_t(\omega_t,p-k) \speq
Z_t(p-k) \;
\frac{\gamma_t}{\pi} \left(
\frac{1}{(\omega_t - \et_{p-k})^2 + \gamma_t^2}
\spp \frac{1}{(\omega_t + \et_{p-k})^2 + \gamma_t^2} \right) \; .
\label{e4.3}
\eeq

After doing the sum over $k^0$ and retaining only terms of leading order
in $g$,
$$
\Gamma_t(\delta E) \speq
g^2 N \, T \; \intk \;
\intw \int^{+\infty}_{-\infty} d\omega_t \;
\frac{\gamma_t}{\omega_t^2 + \gamma_t^2} \; $$
\beq
\left( (1 - (\hat{k} \cdot \hat{p})^2) \rht(\omega,k)
\spm \rhl(\omega,k) \right)
\; \delta(\omega + \omega_t -  E^t_p - \delta E + E^t_{p-k}) \; .
\label{e4.4}
\eeq
The spectral parameter for the hard field has been shifted by
$\omega_t \rightarrow \omega_t + \et_{p-k}$.  With the mass
shells for ultrarelativistic fields, the delta function for
energy conservation fixes the angle between $\hat{p}$ and $\hat{k}$
as $cos\theta = (\omega + \omega_t - \delta E)/k$.

Again I assume that the integrals over $\omega$ and $\omega_t$
decouple.  The integral over $\omega_t$
is peaked about $\omega_t \sim \gamma_t \sim g^2 T$, so there
the effects of $\delta E \neq 0$ must be included.
The integral over the longitudinal spectral density, as in
(\ref{e2.9}), is perfectly well behaved, allowing me to set
$\gamma_t$, $\delta E$, and $m_{mag}$ to zero; this is also
valid for the integral over $\omega \rho_t(\omega,k)$, as
in (\ref{e2.7}).  The integral over $\rho_t(\omega,k)/\omega$,
as in (\ref{e2.10}), is in principal
sensitive to $\omega_f$, $\delta E$, and $\mmag$ when $k \sim \mmag$.
I now make the further assumption, however, that
\beq
\mmag \; \gg \; \gamma_t \; .
\label{e4.40}
\eeq
The limits of integration over $\omega$ run properly over
$\pm k - \omega_t + \delta E$.  When (\ref{e4.40}) holds,
however, even for $k \sim \mmag$ I can neglect the effects of
$\omega_t \sim \delta E \sim \gamma_t$, and just let the
integral over $\omega$ run from $\pm k$, as in the sum rule
of (\ref{e2.8}).

The integral over $\omega_t$ is
\beq
\int^{k+ \delta E}_{- k + \delta E} d\omega_t
\; \frac{\gamma_t}{\omega^2_t + \gamma_t^2}
\speq tan^{-1}\left( \frac{k + \delta E}{\gamma_t} \right)
\spp tan^{-1}\left( \frac{k - \delta E}{\gamma_t} \right) \; .
\label{e4.4a}
\eeq
This is similar to (\ref{e3.8}), except that here $v=1$ and
the shift in the mass shell, $\delta E$, enters.

The integrals
over the soft spectral functions are done using
the sum rules of
(\ref{e2.7}), (\ref{e2.9}), and (\ref{e2.10}).  As in
(\ref{e2.8}) these sum rules are used to
trade an integral from $\pm k$
for terms which involve the right hand side of the sum rule
and pole terms.
In this way
$\Gamma_t = \Gamma_t^{sing}+\Gamma_t^{reg}$ is written as a sum of
a singular term,
$$
\Gamma_{t}^{sing}(\delta E) \speq \frac{g^2 N T}{4 \pi}
\; \int^{\infty}_0 \; k \; dk \;
\left(
\spm \frac{1}{k^2 + m^2_g}
\right.
$$
\beq
\left.
\spp \frac{2}{\pi} \;
\left(
tan^{-1}\left(\frac{k + \delta E}{\gamma_t} \right)
\spp tan^{-1}\left(\frac{k - \delta E}{\gamma_t} \right)
\right)
\; \frac{1}{k^2 + \mmag^2}
\right) \; ,
\label{e4.5}
\eeq
and a regular term,
$$
\Gamma_{t}^{reg} \speq \frac{g^2 N T}{4 \pi}
\int^{\infty}_0 \; k \; dk \left( \frac{1}{k^2 + m^2_g} \right.
\spm \frac{2 Z_t(k)}{\et_k}
$$
\beq \left.
\spm \frac{1}{k^2} \left( 1 \spm 2 Z_t(k) \et_k \right)
\spp \frac{1}{k^2} \left( \frac{3 m^2_g}{k^2 + 3 m^2_g}
\spp \frac{2 k^2 Z_\ell(k)}{\el_k} \right) \right)
\speq 1.09681... \; .
\label{e4.6}
\eeq
So that each integral is finite at
large momentum, a term proportional to
$k/(k^2 + m^2_g)$ is subtracted from
the integrand of the singular term, and
added to that for the regular term.  After
doing so, each term is separately finite and well behaved for both
small and large momentum.
The singular term is sensitive to momenta of order $k \sim g^2 T$,
so there the dependence on $\mmag$, $\gamma_t$, and $\delta E$ must
be retained.  The regular term depends only upon momenta of order
$k \sim m_g$, and so up to corrections of order $g$,
it is a pure number.
This number, $\Gamma_t^{reg} \simeq 1.09681....$,
was determined by numerical integration.

The analytic form of the singular term was
computed in the following manner.
At zero magnetic mass it is easy to show that
\beq
\Gamma_t^{sing}(\delta E, \mmag=0) \speq
\frac{g^2 N T}{8 \pi} \;
ln \left( \frac{m^2_g}{\gamma_t^2 + \delta E^2 } \right)  \; .
\label{e4.6a}
\eeq
Next, I compute the derivative of $\Gamma_t^{sing}$ with respect
to the magnetic mass squared:
\beq
\frac{\partial \Gamma_t^{sing}(\delta E) }{\partial \mmag^2}
\speq \frac{g^2 N T}{8\pi^2} \;
\left( - \; \frac{\gamma_t}{\pi} \right) \int^{+\infty}_{-\infty}
\frac{dk}{(k^2 + \mmag^2)((k + \delta E)^2 + \gamma_t^2)} \; .
\label{e4.6b}
\eeq
This is a standard loop integral in one dimension, and can be
done using the Feynman parametrization for the denominators.
After doing the integrals over $dk$ and the Feynman parameter,
a relatively complicated form for $\partial \Gamma_t^{sing}/
\partial \mmag^2$ results.  Doing the integral over $\mmag$, and
knowing $\Gamma_t^{sing}$ reduces to (\ref{e4.6a}) for $\mmag=0$,
gives a simple result,
\beq
\Gamma_t^{sing}(\delta E) \speq \frac{g^2 N T}{8 \pi} \;
ln \left( \frac{m^2_g}{(\mmag + \gamma_t)^2 + \delta E^2} \right)
\; .
\label{e4.6c}
\eeq

Having computed for real $\delta E$, the analytic continuation of
(\ref{e4.6c}) to complex $\delta E$ is evident.
There are branch points at
\beq
\delta E = \pm i ( \gamma_t + \mmag ) \; ,
\label{e4.6d}
\eeq
which is off the physical sheet.   The damping rate is evaluated at
the pole in the propagator, at $\delta E = i \gamma_t$.
In the limit when $\mmag \gg \gamma$ this pole is
well seperated from the branch point, and the damping
rate is just
\beq
\gamma_t \speq
\Gamma_t^{sing}(i \gamma_t) \spp \Gamma_t^{reg}
\speq \frac{g^2 N T}{8 \pi} \; \left(
\; ln \left( \frac{m_g^2}{\mmag^2 + 2 \mmag \gamma_t} \right) \; + \;
1.09681... \right) \; ,
\label{e4.6e}
\eeq
which is the result quoted in (\ref{e1.1}).

The same manipulations can be carried through for the damping rate
of a (massless) quark field at hard momentum.  I denote
the damping rate of the standard mode, for which the
chirality equals its helicity, as $\gamma_+$.  In the end
the only change is in an overall factor for the
Casimir of the representation:
\beq
\gamma_+ \; = \; \frac{g^2 C_f T}{8 \pi} \; \left(
\; ln \left( \frac{m_g^2}{\mmag^2 + 2 \gamma_+ \mmag} \right) \; + \;
1.09681... \right) \; .
\label{e4.8}
\eeq
These damping rates are the only ones of significance at hard momenta.
At low momenta the gauge field and the quark fields each have
collective modes, the plasmon and plasmino, respectively.  But the
residues of these fields are exponentially small for hard momenta,
so these fields can be neglected.

What of the damping rate of a fast fermion in
hot $QED$, where the magnetic mass vanishes?  In this instance
the above approximations are inconsistent:
the position of the branch point
in the propagator, as in (4.10), coincides with the position of what
is supposed to be a pole.  Thus the spectral density
for the hard fermion is not the pole of a
Breit-Wigner form, but a branch point.
Presumably the damping rate, defined as the
imaginary part of the position of the
branch point in the propagator, is gauge invariant and of
order $e^2 T ln(1/e)$.
The question of gauge dependence is now more involved.
For instance, at zero temperature it is known
that the fermion propagator
has a branch point singularity at the electron mass;
while the position of
the branch point is gauge invariant,
the strength of the singularity is not [\ref{lan}].

The analysis of hot $QCD$ in what is probably the realistic
case of $\mmag \leq \gamma_t$ is even more involved.  Then
the propagator has a pole at $p + i \gamma_t$,
and a branch cut beginning at $p + i (\mmag + \gamma_t)$.
The spectral density for the hard field
must now include the effects of both
the pole and the nearby branch cut.

\section{Damping rates for light fields}

The logarithm in the damping rate of a fast field arises from a
very limited kinematic region: in the one loop diagram, one line
is very near the mass shell, while the other line carries almost
zero momentum.  In this section I analyze the same kinematic
regime for fields moving at momenta comparable to the scale of
the thermal mass.  I just calculate the coefficient of the
logarithm, which is relatively simple to compute.

According to the effective expansion [\ref{bpa}], for soft momenta
the bare propagator is replaced by one which includes the hard
thermal loop, with
the effective plasmon and transverse propagators are
those of (\ref{e2.1}) and (\ref{e2.2}).
For instance, about the mass shell $\omega = \et_p$,
the effective transverse
propagator behaves
\beq
\sDi_t(i \omega,p) \; \sim \; - \; Z^{-1}_t(p) \;
( \omega \spm \et_p ) \; ,
\label{e5.0}
\eeq
where $Z_t(p)$ is the residue.
The leading corrections
to the damping rate are then determined by an effective gluon
self energy, $\sPi^{\mu \nu}$, from which
the transverse term $\sPi_t$ is extracted as usual.
Using (\ref{e5.0}),
the pole of the corrected transverse propagator,
$\sDi_t - \sPi_t$, then determines the damping rate for a
soft transverse gluon, $\gamma_t(p)$ to be
\beq
\gamma_t(p) \; \sim \; Z_t(p) \; Disc \, \sPi_t(i \et_p,p) \; .
\label{e5.1}
\eeq
This is similar to the function introduced in (\ref{e4.1}).
Since I don't compute the constant under the logarithm,
the subtleties of the previous section can be ignored, and it
suffices to evaluate the discontinuity
on the effective mass shell, without including the damping rate.
The damping rates for the plasmon, and the quark modes, are
defined similarly, and given at the end of this section.

{}From (4.21)-(4.25) of ref. [\ref{bpa}], the
effective gluon self energy is a sum of three terms,
\beq
\sPi^{\mu \nu}(P) \; =
\sPit^{\mu \nu}(P) + \sPif^{\mu \nu}(P)
+ \Pi_{gh}^{\mu \nu}(P) \; ,
\label{e5.2}
\eeq
where $\sPit$ is the graph with three gluon vertices,
$$
\sPit^{\mu \nu}(P)  = \; \frac{g^2 N}{2}
\; tr_{soft} \; \sGa^{\sigma \mu \lambda}(-P+K,P,-K)
\; \sD^{\lambda \lambda'}(K)
$$
\beq
\sGa^{\lambda' \nu \sigma'}(-K,P,-P+K) \;
\sD^{\sigma' \sigma}(P-K) \; .
\label{e5.3}
\eeq
and $\sPif$ involves the four gluon vertex,
\beq
\sPif^{\mu \nu}(P) =  - \; \frac{g^2}{2}
\; tr_{soft} \; \sGa^{\mu \nu \lambda \sigma}(P,-P,K,-K)
\sD^{\lambda \sigma}(K) \; .
\label{e5.4}
\eeq
These two terms are the same as in the bare expansion, except that
bare propagators and vertices are everywhere replaced by
effective quantities (the propagators are all in Coulomb gauge).
Lastly, there is the ghost loop,
$\Pi_{gh}$; because there are no hard thermal loops in ghost
amplitudes, this loop equals that in the bare
expansion.  In Coulomb gauge the ghost loop has zero discontinuity,
and so doesn't contribute to the damping rate.
The subscripts on the trace in (\ref{e5.3}) and
(\ref{e5.4}) indicate that the dominant term is given by the integral
over soft momenta.

The effective vertices
which appear in $\sPi$ are nontrivial functions of momenta, and so
the discontinuity of such diagrams is far more complicated than in
the bare expansion.  For example, with a bare vertex the tadpole
diagram has no discontinuity.  In contrast, the discontinuity of
$\sPif$ is nonzero, because the vertex itself has a discontinuity
from Landau damping.

Assume that the leading logarithmic behavior of the damping rate
at nonzero momentum arises from the same kinematic regime
as for a fast field.  Then while the diagram with a four gluon (effective)
vertex, $\sPif$,
does contribute to the discontinuity, it can't
generate a logarithm, since after
cutting through one line and the vertex, there
is no soft line left to integrate over.
Thus the only diagram to contribute is
that with three gluon vertices, $\sPit$.  If the
gluon with momentum $K$ is very soft,
with $K \sim 0$, then the other gluon, with momentum $P-K$, is
very near its mass shell.  (Of course I have to muliply by two,
since the two gluons could be interchanged: $K$ could be near
$P$, so $P-K$ is very soft.)  Now much of
what makes computation
in the effective expansion so involved is the momentum dependence of
the effective vertices.  Under the assumption of very soft $K$, however,
the Ward identities can be used to avoid having to compute any vertices
whatsoever.  Similar Ward identities have also been obtained by
Weldon [\ref{wel}].

The effective three gluon vertex satisfies the Ward identity
\beq
K^\lambda \; \sGa^{\mu \nu \lambda}(P, -P-K,K) \speq
\sDi_{\mu \nu}(P+K) \spm \sDi_{\mu \nu}(P) \; ,
\label{e5.5}
\eeq
Hence as $K \rightarrow 0$,
\beq
\sGa^{\mu \nu \lambda}(P,-P,0) \speq
\frac{\partial}{\partial P^{\lambda}} \; \sDi_{\mu \nu}(P) \; .
\label{e5.6}
\eeq
At small $K$ (\ref{e5.6}) can be used to compute the effective
three gluon vertices which appear in $\sPit$, (\ref{e5.3}).
Even more labor can be saved by recognizing that
only part of (\ref{e5.6}) contributes.

I introduce polarization vectors,
$e^i_a(\hat{p})$.  These are defined as transverse to $\hat{p}$,
\beq
p^i \; e^i_a(\hat{p}) \speq 0 \; ,
\label{e5.7}
\eeq
and an orthonormal set,
\beq
e^i_a(\hat{p}) \; e^i_b(\hat{p}) \speq \delta_{a b} \; .
\label{e5.8}
\eeq
They can be combined to
form a projection operator in $\hat{p}$ as
\beq
P^{ij}(\vec{p}) \speq \delta^{i j} - \hat{p}^i \hat{p}^j \speq
\sum_{a = 1,2} \; e^i_a(\hat{p}) \; e^j_a(\hat{p}) \; .
\label{e5.9}
\eeq
The advantage of using the polarization vectors is that
by sandwiching the effective self energy between the
$e^i$'s, from (\ref{e5.8})
I automatically project onto the transverse part.
Further, for the gluon of momentum $P-K$ in the loop, in
Coulomb gauge the propagator can be approximated as
\beq
\sD^{i j}(P-K) \speq P^{i j}(\vec{p}-\vec{k}) \;
\sDt(P-K) \sim P^{i j}(\vec{p}) \; \sDt(P-K) \; .
\label{e5.10}
\eeq
Then (\ref{e5.9}) is used to write $P^{ij}(\vec{p})$
as a sum over polarization vectors $e^i_a(\hat{p})$.

In this way, {\it all} I need of the Ward identity
in (\ref{e5.6}) are the terms which survive after sandwiching it between
polarization operators.  Using the behavior of the transverse
propagator in (\ref{e5.0}),
\beq
e^i_a(\hat{p}) \; \sGa^{i j k}(P,-P,0) \; e^j_b(\hat{p})
\speq \delta_{a b} \; \hat{p}^k \; Z^{-1}_t(p) \; v_t(p) \; .
\label{e5.11}
\eeq
Here $v_t(p)$ is the group velocity for a transverse gluon of
momentum $p$ on the effective mass shell,
\beq
v_t(p) \speq \frac{\partial \et_p}{\partial p} \; .
\label{e5.12}
\eeq
Of course at hard momenta $v_t(p) \rightarrow 1$.

Computing in the kinematic regime when one transverse gluon is very
soft, and the other gluon is transverse and very near its mass shell,
it is then trivial to mimic the calculations of the previous sections
to extract the leading logarithm in the damping rate.
For the gluon with spectral parameter
$\omega_t$, near its mass shell
I replace the delta functions
for the pole term by smeared Breit-Wigner forms with width $\gamma_t$.
If the soft gluon has spectral parameter $\omega$ the
contribution to the damping rate from a very soft transverse gluon is
$$
\gamma_t(p) \; \sim \;
g^2 N T \;  v^2_t(p) \;
 \int_{k \ll m_g} \frac{d^3 k}{(2 \pi)^3} \;
\intw \int^{+\infty}_{-\infty} d\omega_t \;
\frac{\gamma_t}{\omega_t^2 + \gamma_t^2} \; $$
\beq
\left( (1 - (\hat{k} \cdot \hat{p})^2) \rht(\omega,k) \right)
\; \delta(\omega + \omega_t -  \et_p + \et_{p-k}) \; .
\label{e5.13}
\eeq
This is almost the same integral as for
the damping rate at hard momentum
(\ref{e4.4}), except that each vertex contributes a factor of
the group velocity, and the mass shell is now that for an
effective field.  Expanding the delta function for energy
conservation in small $k$ fixes the angle between $\hat{p}$
and $\hat{k}$ to be $cos\theta = (\omega + \omega_t)/(v_t(p) k)$:
notice the extra factor of one over the group velocity, which is
like the nonrelativistic case in sec. III.  Using the delta
function to integrate over $\theta$ gives one factor of $1/|v_t(p)|$,
so that in all
\beq
\gamma_{t,\ell}(p) \; \sim \;
\frac{g^2 N T}{8 \pi} \; v_{t,\ell}(p) \; ln \left(
\frac{1}{g^2} \right) \spp \ldots \;\;\; .
\label{e5.14}
\eeq
I approximate the argument of the logarithm as
$ln(m_g^2/\mmag^2) \sim ln(1/g^2)$.
As indicated in (\ref{e5.14}), the damping rate of the plasmon can
be computed similarly: it is
given by replacing $v_t(p)$ with the plasmon group velocity,
$v_\ell(p) = \partial \el_p/\partial p$.

For the quark field there are two modes: at positive energy,
the standard mode has
chirality equal to helicity, and while for the plasmino, its
chirality is equal to minus its helicity [\ref{klwf}].
Denoting the mass shells by $E^{\pm}_p$,
in analogy to the result at hard momenta, (\ref{e4.8}),
the damping rate of the quark field is
\beq
\gamma_{\pm}(p) \; \sim \;
\frac{g^2 C_f T}{8 \pi} \; |v_{\pm}(p)| \; ln \left(
\frac{1}{g^2} \right) \spp \ldots \;\;\; ,
\label{e5.15}
\eeq
where $v_{\pm}(p) = \partial E^{\pm}_p/\partial p$ are the group
velocities for the quark modes.  In this instance I explicitly
write the absolute value of the group velocity.  This was not necessary
before, since the mass shells for the transverse, plasmon, and standard
quark modes are each monotonically increasing
in $p$, so the group velocities
are always positive.
The plasmino mass shell, however, decreases from zero
momentum, reaches a minimum at $p = p_c$, and then increases, so
$v_-(p)$ is negative for $p_c > p > 0$.  Similarly,
$\gamma_-(p_c)$ in (\ref{e5.15}) vanishes at $p = p_c$ since
$v_{-}(p_c) = 0$.

The terms in (\ref{e5.14}) and (\ref{e5.15}) apply
for
momenta $p \gg g^2 T$.  For momenta $p \sim T$, all group velocities
$v \rightarrow 1$, and the results of the previous section are recovered.
The restriction
that the momenta be greater than $g^2 T$
is necessary because I assumed that I could approximate
$E_p - E_{p-k} \sim v(p) cos\theta$,
which is incorrect for $p \ll m_g$.
For instance, if one computes as above, but now at exactly zero
momentum, $p=0$, one finds no term as $ln(1/g^2)$ in $\gamma(p)$.
The crossover scale at which a $ln(1/g^2)$ appears
in the damping rate is set by the width of
the spectral density for a field near its mass shell,
which is of order $\sim g^2 T$.  This
restriction is of no concern for the gluon fields, since $v_{t,\ell}(p)
\sim p/m_g$ as $p \rightarrow 0$.  For the quark fields, however,
$|v_{\pm}(p)| \rightarrow 1/3$ as $p \rightarrow 0$, and this caveat is
important: there is no $ln(1/g^2)$ in $\gamma_{\pm}(0)$, as indicated
by the naive extrapolation of (\ref{e5.15}).

This is consistent with explicit calculations at zero momentum.
The damping
rates of both gluons [\ref{bpb}]
and quarks [\ref{bpc},\ref{kkm}] at zero momentum
have been computed:
both results are a pure number times $g^2 T$, with no
terms as $ln(1/g^2)$.  Similarly, the damping rate of the plasmino
at the minimum in its dispersion relation,
$p=p_c$, is surely a pure number times $g^2 T$.

It would take some effort to compute the constants under the logarithm
in the damping rates of (\ref{e5.14}) and (\ref{e5.15}).
Not only does the other diagram, $\sPif$, enter, but for both diagrams
the full form of the vertices are required.  Also, as in sec. IV, the
effects of analytic continuation to the pole in the propagator must be
included.

$\;$\\
\appendix{\noindent{\bf Appendix: $\;\;\;$
The plasmon pressure and sum rules}}
$\;$\\

van Weert and
collaborators [\ref{lw}] raised the following question:
the order $g^3$ term in free energy arises entirely from the
plasmon through the
static electric mass.  The transverse modes do not contribute
because static transverse fields are not screened perturbatively.
On the other hand, for time dependent fields
both the plasmon and transverse fields are screened by thermal masses
$m_g$.  So why don't transverse fields
fields contribute to
the free energy at order $g^3$?
In accord with an analysis by
Toimela [\ref{toi}], in this appendix I show how the sum rules
can be used to demonstrate the cancellation of the transverse
pressure at order $g^3$.

The terms of order $g^3$ in the free energy arise from the summation
of ``ring'' diagrams [\ref{ring}], as the free energy of
effective fields.  For the transverse gluon, per color degree of
freedom the two transverse modes give a free energy
\beq
\sF_t \speq tr \; \left( ln(\sDi_t(K)) \spm  ln(K^2) \right) \; ;
\label{ea.1}
\eeq
as before, $tr$ denotes the integral over the four momentum $K$.
The effective propagator $\sDi_t(K) = K^2 - \dPt(K)$,
where $\dPt(K)$ is the hard thermal loop in the transverse
gluon self energy.  From (\ref{e2.0})
and (\ref{e2.2}), $\dPt(K)$
is $g^2 T^2$ times a function of momentum.  Then it is easy to
compute the derivative of $\sF_t$ with respect to $g^2$:
\beq
\frac{\partial \sF_t}{\partial g^2} \speq
\frac{1}{g^2} \; tr \left( \frac{- \; \dPt(K)}{K^2 - \dPt(K)} \right)
\speq \frac{1}{g^2} \; tr \left( 1 \spm K^2 \; \sDt(K) \right) \; .
\label{ea.2}
\eeq
The sum over $k^0$ is done by using
the spectral representation of the effective propagator,
\beq
tr \left(K^2 \sDt(K) \right) \speq
\intk \; \int^{+\infty}_{-\infty} \; d\omega
\; ( 1 \spp n(\omega)) \; (- \, \omega^2 \spp k^2 ) \rht(\omega,k) \; .
\label{ea.3}
\eeq
The dominant term is given by approximating the Bose--Einstein
statistical distribution
function $n(\omega) \sim T/\omega$,
\beq
tr \left(K^2 \sDt(K) \right) \; \sim \; T \;
\intk \; \int^{+\infty}_{-\infty} \; d\omega
\; \left( - \, \omega  \spp
\frac{k^2}{\omega} \right) \rht(\omega,k) \; .
\label{ea.4}
\eeq
Now use the sum rules of (\ref{e2.7}) and (\ref{e2.10}) to
do the $\omega$ integrals,
\beq
tr \left(K^2 \sDt(K) \right) \; \sim \; T \;
\intk \; \left( 1 \spm \frac{k^2}{k^2 + \mmag^2} \right) \; .
\label{ea.5}
\eeq
Hence if the magnetic mass is ignored, the contribution of transverse
gluons to the pressure cancels identically.  With $\mmag \sim g^2 T$,
up to a possible $\sqrt{ln(1/g^2)}$ [\ref{magc}], the first nonvanishing
contribution to the transverse free energy is of order
$\sF_t \sim g^6 T^4$.  This is where the effects of the magnetic mass
are expected to arise in the free energy [\ref{mag}].

Why did this cancellation occur?  The arguments of van Weert
{\it et al} [\ref{lw}] considered only the contributions of pole
terms to the spectral density.  This is the dominant contribution to the
sum rule of (\ref{e2.7}).  But the effective spectral density of
(\ref{e2.5}) also includes cuts from Landau damping, and it is these
which dominate the sum rule of
(\ref{e2.10}).  For $\mmag = 0$ in $\sF_t$, the contribution of the cuts
identically cancels that of the pole terms.

The analogous contribution of the plasmon to the free energy is,
per color degree of freedom,
\beq
\sF_\ell \speq
\frac{1}{2} \;
tr \; \left( ln(\sDi_\ell(K)) \spm  ln(k^2) \right) \; ,
\label{ea.6}
\eeq
where $\sDi_\ell(K) = k^2 - \dPl(K)$, with $\dPl(K)$ the hard thermal
loop in the plasmon part of the gluon self energy, (\ref{e2.1}).
The derivative of this term with respect to $g^2$ is
\beq
\frac{\partial \sF_\ell}{\partial g^2} \speq
\; \frac{1}{2 g^2} \; tr \left( 1 \spm k^2 \; \sDl(K) \right) \; .
\label{ea.7}
\eeq
Then
\beq
tr( - \, k^2 \sDl(K) ) \sim
T \; \intk \; ( - k^2) \intw \; \rhl(\omega,k) \; .
\label{ea.8}
\eeq
Using the sum rule of (\ref{e2.9}),
\beq
\sF_\ell \; \sim \;
\frac{T}{2} \; \intk \; ln(k^2 \spp  3 \, m^2_g ) \; .
\label{ea.9}
\eeq
Remembering that the static electric mass squared $m^2_{el} = 3 m^2_g$,
(\ref{ea.9}) is exactly equal to the free energy as
computed in the imaginary time formalism, including only the effects
of the term with $k^0 = 0$.  The result for $\sF_\ell$ is a term
of order $g^2$ (this is part of the free energy at two loop order)
plus the term of interest, $= - T m^3_{el}/(12 \pi) \sim g^3 T^4$.
Thus sum rules demonstrate the equivalence
with the results of the imaginary time formalism.

The work of sec. II, and part of sec. IV, was done in collaboration with
E. Braaten, to whom I am indebted.
I also thank R. Baier and A. Weldon for discussions.
This work
was supported in part by the U.S. Department of Energy under
contract DE--AC02--76CH00016.

\noindent{\bf References}
\newcounter{nom}
\begin{list}{[\arabic{nom}]}{\usecounter{nom}}
\item
A. Linde, \phl{93B}{327}{80}.
D. J. Gross, R. D. Pisarski, and L. G. Yaffe, \rmp{53}{43}{81}.
\label{mag}
\item
R. D. Pisarski, \prl{63}{1129}{89}.
\label{rpa}
\item
E. Braaten and R.D. Pisarski,
\prl{64}{1338}{90}; \nup{B337}{569}{90}.
\label{bpa}
\item
E. Braaten and R. D. Pisarski,
\nup{B339}{310}{90};
\prd{42}{2156}{90}.
\label{bpb}
\item
E. Braaten and R. D. Pisarski,
\prd{46}{1829}{92}.
\label{bpc}
\item
J. Frenkel and J. C. Taylor, \nup{B334}{199}{90}.
\label{ft}
\item
J. C. Taylor and S. M. H. Wong, \nup{B346}{115}{90};
E. Braaten and R. D. Pisarski, Phys. Rev. {\bf D45}, R1827 (1992);
J. Frenkel and J. C. Taylor, \nup{B374}{156}{92};
R. Efraty and V. P. Nair, \prl{68}{2891}{92}.
\label{gen}
\item
E. Braaten, R. D. Pisarski, and T. C. Yuan, \prl{64}{2242}{90}.
\label{bpy}
\item
R. Kobes, G. Kunstatter, and K. Mak, \prd{45}{4632}{92}.
\label{kkm}
\item
E. Braaten and T. C. Yuan, \prl{66}{2183}{91};
\label{by}
\item
E. Braaten and M. H. Thoma, \prd{44}{R2625}{91};
{\bf 44}, 1298 (1991).
\label{bt}
\item
R. Baier, H. Nakkagawa, A. Niegawa, and K. Redlich,
\zpc{53}{433}{92}; \prd{45}{4323}{92};
J. Kapusta, P. Lichard, and D. Seibert, \prd{44}{2774}{91}.
\label{pho}
\item
Infrared divergences which signal the appearance of the magnetic mass first
arise in the gluon self energy
at two loop order.  As these are two loop diagrams in an effectively
three dimensional theory, they {\it may} contain a single logarithm:
\beq
\mmag^2 \sim g^4 \; T^2 \; ln\left(\frac{T}{g^2 T}\right) \; ,
\eeq
by which $\mmag \sim g^2 T \sqrt{ln(1/g^2)}$.  Probably, however, gauge
invariance (transversality of the gluon self energy) requires the logarithms
in individual diagrams to cancel in the sum, with $\mmag^2 \sim g^4 T^2$.
\label{magc}
\item
V. V. Lebedev and A. V. Smilga, \anp{202}{229}{90};
\phl{B253}{231}{91}.  Due to an improper introduction of
an angular cutoff, the results for the damping rates in
these works are too small by a factor of $1/3$.
\label{ls}
\item
C. P. Burgess and A. L. Marini, \prd{45}{R17}{92};
A. Rebhan, \prd{46}{482}{92};
T. Altherr, E. Petitgirard, and T. del Rio Gaztelurrutia,
Annecy preprint ENSLAPP-A-378/92 (April, 1992).
\label{fast}
\item
R. Baier, H. Nakkagawa, and A. Niegawa, Ecole Polytechnique
preprint A157-0392 (March, 1992); Osaka City University preprint
OCU-PHYS-145 (September, 1992).
\label{bnn}
\item
R. Kobes, G. Kunstatter, and A. Rebhan, \prl{64}{2992}{90};
\nup{B355}{1}{91}.
\label{kkr}
\item
R. Baier, G. Kunstatter, and D. Schiff, \prd{45}{4381}{92};
Bielefeld preprint BI-TP 92/19 (June, 1992).
\label{bks}
\item
A. Rebhan, CERN preprint CERN-TH-6434-92.
\label{reb}
\item
H. Nakkagawa, A. Niegawa, and B. Pire,
Ecole Polytechnique preprints A156-0292 (February, 1992) and
A195-0992 (September, 1992).
\label{nnp}
\item
V. P. Silin, Sov. Phys. J.E.T.P. {\bf 11}, 1136 (1960);
J.-P. Blaizot and E. Iancu, Saclay preprint (May, 1992).
\label{kin}
\item
V. V. Klimov, Sov. J. Nucl. Phys. {\bf 33}, 934 (1981);
Sov. Phys. J.E.T.P.
{\bf 55}, 199 (1982);
H. A. Weldon, \prd{26}{1394}{82}.
\label{klwg}
\item
R. D. Pisarski, Physica {\bf A 158}, 146 (1989).
\label{rpb}
\item
H. A. Weldon, \prd{40}{2410}{89};
H. Schulz, Hannover preprint ITP-UH 5/92.
\label{wel}
\item
Relativistic Quantum Theory, E. M. Lifshitz and L. P. Pitaevskii,
(Pergamon Press, London, 1977); sec. 116, especially (116.18).
\label{lan}
\item
V. V. Klimov, Sov. Phys. J.E.T.P. {\bf 55}, 199 (1982);
H. A. Weldon, \prd{26}{1394}{82};
R. D. Pisarski, \nup{A498}{423c}{89}.
\label{klwf}
\item
N. P. Landsman and Ch. G. van Weert,
Phys. Rep. {\bf 145}, 141 (1987).
\label{lw}
\item
T. Toimela, \phl{B176}{463}{86}; {\bf 192}, 427 (1987).
\label{toi}
\item
I. A. Akhiezer and S. V. Peletminsky, Sov. Phys. JETP {\bf 11},
1316 (1960); J. I. Kapusta, \nup{B148}{461}{79}.
\label{ring}
\end{list}
\end{document}